\documentclass{ws-ijmpa}
\usepackage[super,compress]{cite}

\usepackage{amsmath,amssymb,amsfonts}
\usepackage{graphicx}
\usepackage{color}

\newcommand{\foom}[1]{\protect\footnotemark[#1]}

\def\inch{\hspace*{1in}}

\def\Jl#1#2{{\it #1\/} {\bf #2},\ }

\def\ApJ#1 {\Jl{Astroph. J.}{#1}}
\def\CQG#1 {\Jl{Class. Quantum Grav.}{#1}}
\def\DAN#1 {\Jl{Dokl. AN SSSR}{#1}}
\def\GC#1 {\Jl{Grav. Cosmol.}{#1}}
\def\GRG#1 {\Jl{Gen. Rel. Grav.}{#1}}
\def\JETF#1 {\Jl{Zh. Eksp. Teor. Fiz.}{#1}}
\def\JETP#1 {\Jl{Sov. Phys. JETP}{#1}}
\def\JHEP#1 {\Jl{JHEP}{#1}}
\def\JMP#1 {\Jl{J. Math. Phys.}{#1}}
\def\NCB#1 {\Jl{Nuovo Cim. B}{#1}}
\def\NPB#1 {\Jl{Nucl. Phys. B}{#1}}
\def\NP#1 {\Jl{Nucl. Phys.}{#1}}
\def\PLA#1 {\Jl{Phys. Lett. A}{#1}}
\def\PLB#1 {\Jl{Phys. Lett. B}{#1}}
\def\PRD#1 {\Jl{Phys. Rev. D}{#1}}
\def\PRL#1 {\Jl{Phys. Rev. Lett.}{#1}}

\def\lal{&&{}}
\def\eq{Eq.\,}

\def\beq{\begin{equation}}
\def\eeq{\end{equation}}
\def\bear{\begin{eqnarray}}
\def\bearr{\begin{eqnarray} \lal}
\def\ear{\end{eqnarray}}
\def\earn{\nonumber \end{eqnarray}}

\def\nnn{\nonumber\\ \lal }

\def\e{{\,\rm e}}

\def\eps{\varepsilon}

\def\rf{\eqref}
\def\mn{_{\mu\nu}}
\def\MN{^{\mu\nu}}
\def\mN{_\mu^\nu}

\def\kappa{\varkappa}

\def\sph{spherically symmetric}

\def\bh{black hole}

\def\Swz{Schwarz\-schild}

\begin{document}

\markboth{K.A. Bronnikov, S.V. Bolokhov and M.V. Skvortsova}
{A possible semiclassical bounce instead of a Schwarzschild singularity}


\title{A POSSIBLE SEMICLASSICAL BOUNCE INSTEAD OF A SCHWARZSCHILD SINGULARITY}

\author{K. A. BRONNIKOV\foom 1}

\address{VNIIMS, 	Ozyornaya 46, Moscow 119361, Russia;\\
	Institute of Gravitation and Cosmology, Peoples' Friendship University of Russia\\
	(RUDN University), ul. Miklukho-Maklaya 6, Moscow 117198, Russia;\\
	National Research Nuclear University ``MEPhI''
		(Moscow Engineering Physics Institute),\\
		Kashirskoe sh. 31, Moscow 115409, Russia\\
		kb20@yandex.ru		}
		
\author{S.V. BOLOKHOV}		

\address{Institute of Gravitation and Cosmology, Peoples' Friendship University of Russia\\
	(RUDN University), ul. Miklukho-Maklaya 6, Moscow 117198, Russia;\\
		boloh@rambler.ru        }
		
\author{M.V. SKVORTSOVA}	
\address{Peoples' Friendship University of Russia	(RUDN University),\\ 
		ul. Miklukho-Maklaya 6, Moscow 117198, Russia;\\
		milenas577@mail.ru        }

\maketitle

\begin{history}
\received{Day Month Year}
\revised{Day Month Year}
\end{history}

\begin{abstract}
  We have previously shown that the singularity in a Schwarzschild black hole of stellar or larger mass
  may be avoided in a semiclassical manner by using as a source of gravity the stress-energy tensor (SET)
  corresponding to vacuum polarization of quantum fields, with a minimum spherical radius a few orders 
  of magnitude larger than the Planck length.  In this note we estimate the nonlocal contribution to the 
  total SET due to particle creation from vacuum. We show that this contribution is negligibly small
  as compared to vacuum polarization and does not affect the previously suggested scenario.  
\end{abstract}


\ccode{PACS numbers: 97.60.Lf, 04.70.-s, 04.70.Dy
}

  In our recent paper \cite{we18} we proposed a simple semiclassical model of the interior of a 
  \Swz\ \bh\ (BH) showing that  the \Swz\ singularity may in principle be avoided by including into 
  consideration the stress-energy tensor (SET) describing vacuum polarization of quantum fields 
  in the form of the tensors $^{(1)}H\mN$ and  $^{(2)}H\mN$ obtained by variation of curvature-quadratic 
  terms $R^2$ and $R\mn R\MN$ in the effective action. Under certain conditions for the indefinite
  constant factors before these tensors, it has turned out to be possible to obtain a regular bounce 
  instead of the \Swz\ singularity. In this scenario, the spherical radius $r$ has a regular minimum 
  (instead of zero), while the longitudinal scale experiences a regular maximum (instead of infinity) 
  in the corresponding Kantowski-Sachs metric. The free parameters of the model may be chosen 
  in such a way that the curvature scale remains a few orders of magnitude below the Planck scale 
  (in particular, on the GUT scale), so that the bounce is implemented on a semiclassical level without 
  invoking any (so far vague) quantum gravity effects.
  
  The tensors $^{(1)}H\mN$ and  $^{(2)}H\mN$ represent a local part of the entire quantum-field 
  contribution to the SET of matter. There is also a nonlocal part, depending on the whole history and 
  mainly represented by particle production from vacuum. In this short note we try to estimate the order 
  of magnitude of this nonlocal contribution to the effective SET due to processes in the vicinity of a 
  bounce.

  We consider a model of a bouncing \sph\ BH interior \cite{we18} (T-region)
  described by the Kantowski-Sachs metric
\beq
	ds^2 = \e^{2\alpha} d\eta^2 - \e^{2\gamma}dx^2 - \mu^2\e^{2\beta}d\Omega^2,
\eeq
  where the time coordinate $\eta$ is the so-called ``conformal time,'' specified by the condition 
  $3\alpha(\eta) = 2\beta(\eta) + \gamma(\eta)$ convenient for considering quantum fields. We assume 
  that the BH has a stellar (or larger) mass $m_{\rm Sch}$, and $\mu=2m_{\rm Sch}\gtrsim 10^5$\,cm
  is the corresponding gravitational radius (in units with $c=G=1$). On the other hand, for the bounce
  (say, at $\eta=0$),  to provide a semiclassical regime, we assume that the minimum radius 
  $r_0 = \mu \e^{\beta(0)}$ is  $\sim 10^5 l_{\rm Pl}\sim 10^{-28}$\,cm. Introducing the small parameter 
  $\eps =r_0/\mu \lesssim 10^{-33}$, for times close to the bounce we can write
\beq  				\label{exps}
	\e^{2\alpha} = \eps(1+a \eta^2),  	\qquad
	\e^{2\beta} = \eps^2(1+b \eta^2),	\qquad
	\e^{2\gamma} = \eps^{-1}(1+c \eta^2).
\eeq
  with $3a = 2b + c$ by the definition of $\eta$, $b > 0$ since $\e^\beta$ has a minimum, and $c < 0$ 
  since $\e^\gamma$ has a maximum at $\eta =0$. The powers of $\eps$ correspond to magnitudes 
  of the metric coefficients at approach to a would-be \Swz\ singularity.

  Consider the standard Fourier expansion for a quantum scalar field 
\beq
	\Phi={\cal N} \e^{-\alpha}\int dk \sum_{lm}\e^{-ikx} Y_{lm}(\theta, \varphi)g_{klm}(\eta) c^{+}_{klm}+{\sf h.c.},
\eeq
  where ${\cal N} $ is a normalization factor, $ c^{+}_{klm}$ is a creation operator, $Y_{lm}$ are 
  spherical functions, and each mode function $g_{klm}(\eta)\equiv g$ obeys the equation obtained
  by separation of variables in the original Klein--Gordon-type equation:
\beq
		\ddot{g}+\Omega^2 g=0,
\eeq
  where the dot stands for $d/d\eta$, and $\Omega$ is the effective frequency:
\beq 		\label{Omega}
		\Omega^2=k^2\e^{2(\alpha-\gamma)}+\frac{l(l+1)+2\xi}{\mu^2}
		\e^{2(\alpha-\beta)}+M^2\e^{2\alpha}+\frac{2\xi (\dot\beta-\dot\gamma)^2}{3}+
		(6\xi-1)(\ddot\alpha+\dot\alpha^2).
\eeq
   At bounce time $\eta =0$ we have, by normalization, $|g| \sim \Omega^{-1/2}$ and 
\beq		\label{Omega0}
	\Omega^2(0) = k^2\eps^2+\frac{l(l+1)+2\xi}{\mu^2\eps} + M^2\eps + (6\xi-1)a.
\eeq

  Now, for estimation purposes,  we make a natural assumption, confirmed by experience,\cite{BDa, GMM} 
  that particle creation occurs most intensively with energies close to the curvature scale $\sim r_0^{-1}$.  
  This energy is of the order of the frequency $\bar\Omega(\tau)$ in terms of the proper time $\tau$ 
  related to the conformal time by $d\tau=\e^\alpha d\eta$. Thus our assumption means 
  $\bar\Omega \sim 1/r_0$. Since $\e^\alpha \sim \sqrt{\eps}$, one has $\tau\sim\sqrt{\eps}\eta$, and 
  from the relation $\Omega\eta = \bar\Omega\tau$ we obtain $\bar\Omega=\Omega/\sqrt{\eps}$, so that
\beq 		\label{Omegabar0}
		\bar\Omega^2(0)=k^2\eps+\frac{l(l+1)+2\xi}{\mu^2\eps^2} + M^2+\frac{(6\xi-1)a}{\eps}.
\eeq
  It is of interest which values of the parameters involved make contributions to 
  $\bar\Omega^2$  of the order $\sim r_0^{-2} = (\mu\eps)^{-2}$. They are:
\beq                           \label{orders}
	k \sim \frac{1}{\mu^{2}\eps^{3}}\sim10^{45}\,{\rm cm}^{-1}\sim 10^{12}\,m_{\rm Pl}; \quad
	 l,\xi\sim 1; \quad     M\sim \frac 1 {r_0};  \quad 
	 a = \ddot\alpha(0) \sim \frac {\eps}{ r_0^2}.
\eeq
  Notably, momenta $k$ much larger than Planckian look meaningless, and we conclude that 
  reasonable (sub-Planckian) values of $k$ do not appreciably contribute to $\bar\Omega$. 

  The result $a\sim \eps/r_0^2$ can be obtained independently from the relations
  $e^{2\alpha}=\eps^{-1}(1+\bar a \tau^2)=\eps^{-1}(1+a\eta^2)$, $\bar a\sim r_0^{-2}$, 
  $\tau\sim\sqrt{\eps}\eta$. A similar analysis also gives $b, c\sim \eps/r_0^2$.
  Furthermore, at small $\eta$ we can suppose
\beq 			\label{OmegaTaylor}
	\Omega\approx B + C \eta^2, \quad\  \text{where}\ \  
	B=\Omega(0)\sim\sqrt{\eps}r_0^{-1},  \quad C/B\sim (a,b,c)\sim \eps /r_0^2.
\eeq

  To estimate the energy density of created particles, one can use standard technique of Bogoliubov
  coefficients. For the case of a bounce-type metric, the crucial coefficient $\beta_{kl}$ can be 
  calculated with necessary accuracy using the formulas \cite{Branden} 
\beq                  \label{II}
	\beta_{kl}=\sqrt{\frac{I^{-}}{I^{+}}}\sinh\sqrt{I^{-}I^{+}},
	\quad I^{\pm}\equiv\int_{\eta_1}^{\eta}g^{\pm}(\bar\eta)d\bar\eta,
	\quad g^{\pm}\equiv\frac{\dot\Omega}{2\Omega}\exp\left(\pm 2i\int_{\eta_1}^\eta 
		\Omega(\bar{\eta})d\bar{\eta}\right),
\eeq
  where $\eta_1$ is the initial time at which, by assumption, $\beta_{kl}=0$ (a vacuum 
  state of the field, absence of particles). Using \eq\eqref{OmegaTaylor} and assuming 
  $B\eta\lesssim O(1)$ (that is, $\eta$ is not too far both from zero and from $\eta_1)$,  we have
\bearr
	\int_{\eta_1}^\eta \Omega(\bar{\eta})d\bar{\eta} \approx 
			B\bar\eta+\frac{1}{3}C\bar\eta^3\Bigr|^\eta_{\eta_1}\approx  B (\eta-\eta_1),
\\ \lal
	g^{\pm}(\eta) \approx \frac{C\eta}{B}\e^{\pm 2i B(\eta-\eta_1)}
			\sim \frac{\eps\eta}{r_0^2}\e^{\pm 2i B(\eta-\eta_1)}.
\ear
  Now we can calculate the integrals $I^\pm$ involved in \rf{II} close to bounce ($\eta=0$):
\bearr
	I^{\pm}(\eta)\Bigr|_{\eta\to0}\sim \frac{\eps}{r_0^2}\int_{\eta_1}^0\eta d\eta \e^{\pm 2i  
		B(\eta-\eta_1)}=\frac{\eps}{r_0^2}\e^{\mp2iB\eta_1}
			\left[\frac{\e^{\pm2iB\eta}}{4B^2}(1\mp2iB\eta) \right]^0_{\eta_1}
\nnn
		\inch =\frac{1}{4}\left[\e^{\mp2iB\eta_1}-1\pm2iB\eta_1\right]\approx-\frac{1}{2}B^2\eta_1^2.
\ear
  Then, assuming $B\eta_1\lesssim O(1)$, we have
\beq
\beta_{kl}\sim I^{-}\sim-\frac{1}{2}B^2\eta_1^2, \qquad
|\beta^2_{kl}|\sim\frac{1}{4}B^4\eta_1^4\lesssim O(1).
\eeq 
  Thus the energy density of created particles is
\beq             \label{rho}
		\rho_{\rm nonloc}=\langle T^0_0\rangle \sim \frac{1}{8\pi}
	\int dk \sum_l(2l+1)\frac{e^{-4\alpha}}{\mu^2}\Omega |\beta_{kl}|^2
		\sim\frac{10^5\sqrt{\eps}}{r_0^4}\sim \frac{10^{-11}}{r_0^4},
\eeq
  where we have used the following approximate orders of magnitude for each factor in \rf{rho} in 
  agreement with \rf{orders}:
  (i) $\int dk\sim 2m_{\rm Pl} = {10^5}/{r_0}$ since we integrate from $-m_{\rm Pl}$ to $+m_{\rm Pl}$; 
  (ii) $\sum_l (2l+1) \sim 10^2$, including a few low multipolarities; 
  (iii)  $e^{-4\alpha}/\mu^2 \sim {1}/{r_0^2}$; (iv) $\Omega \sim \sqrt{\eps}/r_0$; 
  (v) $|\beta_{kl}|^2 \sim 1$ as a rough upper bound.
  
  Comparing the estimate \rf{rho} with the local energy density contribution \cite{we18} of the order 
  $\rho_{\rm loc}\sim 10^{10}r_0^{-4}$, we see that  $\rho_{\rm nonloc}/\rho_{\rm loc}\sim 10^{-21}$,
  and this value is still smaller for BH masses larger than that of the Sun.
  Thus the nonlocal contribution due to particle creation turns out to be negligibly small in the 
  semiclassical bounce regime, and this estimate can hardly change too strongly if this contribution 
  is calculated more precisely and if more physical fields of different spins are included. 

   As mentioned in Ref. \refcite{we18}, our toy model describes a possible geometry that may be 
   seen by an observer falling into a sufficiently large BH for which Hawking radiation is negligible 
   due to its extremely low temperature. It would be, however, of great interest to study how this
   model will change if Hawking radiation is included, as well as to study similar models for BHs 
   with charge and spin.  

\section*{Acknowledgments}

  This publication was supported by the RUDN University program 5-100. 
  The work is also supported by RFBR Grant No. 19-02-00346. The work of KB was also partly 
  performed within the framework of the Center FRPP supported by MEPhI Academic Excellence 
  Project (contract No. 02.a03. 21.0005, 27.08.2013).

\end{document}